\documentclass[twocolumn]{aastex62}

\newcommand{\RNum}[1]{\uppercase\expandafter{\romannumeral #1\relax}}
\newcommand{\bef}{\begin{figure}}      
\newcommand{\eef}{\end{figure}}      
\newcommand{\bea}{\begin{eqnarray}}    
\newcommand{\eea}{\end{eqnarray}}      
\newcommand{\be}{\begin{equation}}      
\newcommand{\ee}{\end{equation}}  
\newcommand\HI{$\textrm{H}\scriptstyle\mathrm{I}$}

\graphicspath{{./}{figures/}}

\received{???}
\revised{???}
\accepted{???}
\shorttitle{Mass models of the Milky Way from the GAIA DR3 data-set}
\shortauthors{Sylos Labini et al.}
\usepackage{amsmath}

\begin{document}

\title{Generalized rotation curves of the Milky Way from the GAIA DR3 data-set: constraints on mass models}
\author{Francesco  Sylos Labini}
\affil{Centro  Ricerche Enrico Fermi, Via Pansiperna 89a, 00184 Rome, Italy}
\affil{Istituto Nazionale Fisica Nucleare, Unit\`a Roma 1, Dipartimento di Fisica, Universit\'a di Roma ``Sapienza'', 00185 Rome, Italy}

\correspondingauthor{FSL}
\email{sylos@cref.it}

\begin{abstract}
The circular velocity curve traced by stars provides a direct means of investigating the potential and mass distribution of the Milky Way. Recent measurements of the Galaxy's rotation curve have revealed a significant decrease in velocity for galactic radii larger than approximately 15 kpc. While these determinations have primarily focused on the Galactic plane, the Gaia DR3 data also offer information about off-plane velocity components.
By assuming the Milky Way is in a state of Jeans equilibrium, we derived the generalized rotation curve for radial distances spanning from 8.5 kpc to 25 kpc and vertical heights ranging from -2 kpc to 2 kpc. These measurements were employed to constrain the matter distribution using two distinct mass models. The first is the canonical NFW halo model, while the second, the dark matter disk (DMD) model, posits that dark matter is confined to the Galactic plane and follows the distribution of neutral hydrogen.
The best-fitting NFW model yields a virial mass of $M_{\text{vir}} = (6.5 \pm 0.5) \times 10^{11} M_\odot$, whereas the DMD model indicates a total mass of $M_{\text{DMD}} = (1.7 \pm 0.2) \times 10^{11} M_\odot$. 
Our findings indicate that the DMD model generally shows a better fit to both the on-plane and off-plane behaviors at large radial distances of the generalized rotation curves when compared to the NFW model. We emphasize that studying the generalized rotation curves at different vertical heights has the potential to provide better constraints on the geometrical properties of the dark matter distribution.
\end{abstract}

\section{Introduction}

The circular velocity curve of the Milky Way (MW) has been measured using various tracers and methods, depending on the range of galactocentric radii ($R$) being considered (see, e.g., \cite{Bhattacharjee_etal_2014, Sofue_2020} for recent reviews). For $R$ values smaller than the solar radius ($R_\odot \sim 8$ kpc), the rotation curve can be derived using the tangent-point method that involves measuring the radio emission from  \HI{}  and CO lines of the interstellar medium \citep{Gunn_etal_1979, Fich_etal_1989, Levine_etal_2008, Sofue_etal_2009}. For $R > R_\odot$, specific samples of stars with measurable distances, proper motions, and/or line-of-sight velocities have been used to constrain the rotation curve. These include classical cepheids \citep{Pont_etal_1997}, red clump giants \citep{Bovy_etal_2012,LopezCorredoira_2014,Huang_etal_2016}, RR Lyrae stars \citep{Ablimit+zhao_2017, Wegg_etal_2019}, and blue horizontal branch stars \citep{Xue_etal_2009, Kafle_etal_2012}. However, these stellar standard candles are often rare or not bright enough to be observable at large distances, and uncertainties in distance estimates can introduce significant errors in the analysis of the circular velocity curve. Furthermore, the full three-dimensional velocity information of the tracers is generally not available, so the circular velocity has to be estimated using only the measured line-of-sight velocity and position. This estimation necessarily assumes that the emitter is in a perfectly circular orbit.

To accurately determine the MW's rotation curve without relying on key assumptions about the kinematic of emitters, precise measurements of the Galactocentric radius, tangential velocity, and radial velocity for each star are necessary, including the uncertainties in position and velocity in all three spatial dimensions. These measurements enable the computation of the various terms in the Jeans equation that, {  assuming the Galaxy is close to a self-gravitating steady state},  provides a connection between the derivatives of the gravitational potential, the mass density and the  moments of the velocity components  (see, e.g., \cite{Binney+Tremaine_2008}).

The Gaia mission has recently provided a large sample of stars with high precision parallax and proper motion measurements  \citep{Gaia_2016, Gaia_2018, Gaia_2021}. Additionally, the third data release of the Gaia mission (DR3) has significantly increased the catalog of stars' line-of-sight velocities, with over 30 million stars included \citep{Katz_etal_2022}. The availability of all six dimensions for a large sample of stars in the MW marks a new phase in determining its rotation curve and of a number of other kinematic properties \citep{Antoja_etal_2021,Drimmel_etal_2022,Katz_etal_2022}.

Recently, several  research groups have utilized the Gaia datasets to determine the MW's rotation curve, employing different samples of stars \citep{Eilers_etal_2019, Mroz_etal_2019,Chrobakova_etal_2020,Wang_etal_2023, Ou_etal_2024}. The measurements by \cite{Eilers_etal_2019} and \cite{Mroz_etal_2019} are based  on samples of red giant stars and Cepheids, respectively. \cite{Wang_etal_2023} obtained the rotation curve by applying a statistical deconvolution of parallax errors using Lucy's inversion method (LIM) \citep{Lucy_1977} to the full sample of Gaia DR3 sources. This method was firstly applied by \cite{Lopez-Corredoira_Sylos-Labini_2019} to the Gaia DR2 data; the rotation curve was then determined by \cite{Chrobakova_etal_2020}. Finally, \cite{Ou_etal_2024} presented an updated circular velocity curve obtained in a a sample of red giant stars, that was larger than the one considered by \cite{Eilers_etal_2019}.

The new measurements  indicate that the MW's rotation curve $v_c(R)$ is not flat but exhibits a gradual decline, transitioning from approximately 230 km s$^{-1}$ at 5 kpc to around 170 km s$^{-1}$ at 28 kpc. Specifically, both  \cite{Wang_etal_2023} and \cite{Ou_etal_2024} found that the circular velocity curve declines at a faster rate for large galactic radii ($R > 20$ kpc) compared to inner galactic radii. A similar trend, albeit not definitive, was observed in \cite{Eilers_etal_2019}, while \cite{Mroz_etal_2019} found a slower decline at smaller radial distances (i.e., $R<15$ kpc). In their study, \cite{Yongjun_etal_2023} compared different estimates of the MW's rotation curve and conducted a robust assessment of the systematic uncertainties. They confirmed a significant decrease in velocity between 19.5 and 26.5 kpc, amounting to approximately 30 km s$^{-1}$. They have interpreted this observation as indicating a Keplerian decline in rotation, which initiates at a distance of 19 kpc from the Galaxy center and extends to 26.5 kpc. Moreover, they rejected the hypothesis of a flat rotation with a statistical significance of 3$\sigma$.

The fact that $v_c(R)$ decreases with increasing radius implies that the mass estimated in a given galactic mass model should be lower than that for a flat rotation curve.
{  However, a certain amount of dark matter (DM) is required, as the stellar components alone are insufficient to account for the observed velocity profile \citep{Eilers_etal_2019, SylosLabini_etal_2023_MW, Ou_etal_2024}, unless a modified gravity model is invoked  \citep{Milgrom_1983,McGaugh_etal_2016,Lopez-Corredoira+Betancort-Rijo_2021}.} 
In this regard, the best fit obtained with the canonical Navarro-Frank-White (NFW) halo model \citep{Navarro_etal_1997}, that assume DM distributed approximately spherically, yields a virial mass of $M_{\text{vir}} = (6.5 \pm 0.5) \times 10^{11} M_\odot$ and a virial radius of $R_{\text{vir}} = (180 \pm 3)$ kpc \citep{SylosLabini_etal_2023_MW}. These values are approximately 20\% smaller than the estimation for this same model provided by \cite{Eilers_etal_2019}, who had data only up to 25 kpc. In addition, both datasets yield lower mass estimates compared to several previous studies that used an approximately constant $v_c(R)$ (see, e.g., \cite{Bovy_etal_2012,Eadie+Harris_2016,Eadie_etal_2018}). 

More recently, \cite{Ou_etal_2024}  fitted to their determination of the rotation curve  two different  mass models for the DM halo: a generalized NFW (gNFW) profile and an Einasto profile \citep{Einasto_1965,Retana-Montenegro_etal_2012}. Both models introduce an additional free parameter compared to the standard NFW so that they are more suitable to fit a declining rotation curve. In particular, the gNFW profile incorporates a parameter that modulates the inner and outer asymptotic power-law slope of the standard NFW profile. The halo virial masses obtained for this model is $M_{\text{vir}} = (5.17 \pm 0.1) \times 10^{11} M_\odot$,  similar to  the value found by \cite{SylosLabini_etal_2023_MW} in the NFW case. The Einasto model is described by a stretched exponential profile, with the exponent playing the role of the additional free parameters beyond the total mass and characteristic length scale;  in this model the density profile decays faster than in the NFW case,  and the best fit  gives  $M_{\text{vir}} = (1.5 \pm 0.04) \times 10^{11} M_\odot$ that is significantly lower than previous estimations. 

Finally \cite{SylosLabini_etal_2023_MW}  found a MW's mass of $ (1.6 \pm 0.5) \times 10^{11} M_\odot$  for a    model that assumes DM to be confined in  the Galactic disc: this is named as the dark matter disk (DMD) model. The motivation for considering this model stems from the "Bosma effect" \citep{Bosma_1978,Bosma_1981}, which is an observation in external disc galaxies suggesting a correlation between DM and neutral hydrogen (\HI{}). Indeed, there is substantial observational evidence indicating that rotation curves of external disc galaxies, particularly at larger radii, exhibit a rescaled version of those derived from the \HI{} distribution \citep{Sancisi_1999,Hoekstra_etal_2001,Hessman+Ziebart_2011, Swaters_etal_2012,SylosLabini_etal_2024_Mass}. Even in the case of the MW, it is  possible to fit the rotation curve by positing that the distribution of  \HI{} serves as a proxy for DM \citep{SylosLabini_etal_2023_MW}.

The studies mentioned above primarily focused on determining the rotation curve within the plane of the galaxy. Recently, some  datasets have allowed for exploration of off-plane regions and the investigation of the vertical dynamics of the MW. These analyses have primarily been undertaken to differentiate between two main hypotheses: the existence of an approximately spherical DM halo, such as the NFW model, and fitting the rotation curve by assuming modified Newtonian dynamics  (MOND --- see, e.g., \cite{Milgrom_1983,McGaugh_etal_2016,Lopez-Corredoira+Betancort-Rijo_2021}). The latter hypothesis proposes that the flat rotation curves observed in the outer regions of disc galaxies are not the result of a massive DM halo, but rather indicative of MOND.  \cite{Nipoti_etal_2007}  pointed out that that while these models can produce nearly identical rotation curves within the disc, they exhibit distinctive differences in terms of vertical dynamics.
In a recent study by \cite{Zhu_etal_2023}, the complete form of the Jeans equations was employed to differentiate between various mass models of the MW. This approach has been previously discussed in studies such as \cite{Kipper_etal_2016} and references therein.
The authors utilized two independent Jeans equations, namely the radial and vertical directional equations, as discriminators to assess the consistency between gravitational potential models and kinematic data. Under the assumptions of a stationary system with zero average radial and vertical velocity, and axi-symmetry resulting in zero cross-terms of the velocity dispersion tensor, the relevant kinematic quantities entering the Jeans equations were identified.%

To estimate these kinematic quantities, the study analyzed velocity data from the LAMOST and Gaia red clump sample compiled by \cite{Huang_etal_2020}. This sample consisted of approximately 137,000 red clump stars within a range of 4 kpc to 16 kpc in galactocentric distance and within a height of 4 kpc in vertical distance.
 The results of these analyses revealed that these models were equally consistent with the data at almost all spatial locations within the analyzed range.

In the present study, we employ {  both the original  Gaia DR3 data and the extendend} kinematic maps obtained from Gaia DR3 data, which were analyzed using the LIM technique \citep{Lopez-Corredoira_Sylos-Labini_2019,Wang_etal_2023}, to explore the off-plane dynamics. Indeed, these 
{  data}  enable us to derive velocity moments both within and outside the galactic plane. {  In particular, the direct Gaia DR3 data have an upper limit for the radial distance of $R<14$ kpc, whereas the extended kinematic maps cover a radial range between 12 kpc and 22 kpc; in both cases the vertical range extends from -2 kpc to 2 kpc.} Assuming the Galaxy is in a steady state {  under its self-gravity} and utilizing the complete set of Jeans equations, we compare the generalized rotation curves in the plane and off the plane to the predictions of the standard NFW halo model and the DMD  model. While we do not perform a complete $\chi^2$ minimization over the free parameters  of these  models for the off-plane case, we can still obtain valuable insights regarding the compatibility of the data with the two mass distributions considered.

The paper is structured as follows: in Section \ref{jeans_sect}, we recall the basic elements of the Jeans equations that will be used in our study stressing the underlying approximations. In particular, we emphasize the features of the multicomponent mass models. In Section \ref{gaiadr3}, we discuss the determination of the kinematic quantities from the Gaia DR3 data and make the best fits of the mass models that we have considered. Finally, in Section \ref{conclusion}, we present our conclusions.


\section{Jeans equations and mass models}
\label{jeans_sect}

\subsection{Jeans equations}
\label{jeans_equations} 
The problem at hand involves measuring kinematic quantities and obtaining information about the density distribution of luminous components (such as stars and gas, i.e. the baryonic components) in order to constrain the total mass  based on a given mass model. To achieve this objective, the {  self-consistent} Jeans-Poisson system of equations is employed, assuming that the Galaxy is in a steady state so that all time derivatives are equal to zero  \citep{Binney+Tremaine_2008}. The Jeans equation provides a reasonable approximation for astrophysical systems like galaxies, as the timescale for collisions between stars is significantly longer compared to the crossing  timescale $\tau \approx \sqrt{G\rho}^{-1}$,  so that the fundamental dynamics is that of a collisionless system. This treatment neglects any time-dependent physical processes. 

Let us suppose that the system under consideration is axi-symmetric. The first Jeans equation relates velocity moments to the radial acceleration   
\bea
\label{eq:jeans1}
\frac{\partial \rho \overline{v^2_R}} { \partial R} + 
\frac{ \partial \rho \overline{v_R v_z }}{ \partial z} + 
\rho \left( \frac{\overline{v^2_R} - \overline{v^2_\theta}}{R} + 
\frac{ \partial \Phi}{ \partial R} \right) =0 \;.
\eea 
The second  Jeans equation relates velocity moments to the vertical acceleration 
\bea
\label{eq:jeans2}
\frac{\partial \Phi} { \partial z} = 
- \frac{\overline{v_R v_z}} { R} 
- \frac{1} { \rho} \frac{\partial \rho\overline{v_R v_z}} { \partial R} 
- \frac{1} { \rho} \frac{\partial \rho\overline{v_z^2}} { \partial z} \;.
\eea 
If the gravitational potential is generated only by the mass density $\rho$ then this is given by  given by the Poisson's equation
\be
\label{eq:poisson}
\nabla^2 \Phi = 4 \pi G \rho \;. 
\ee
{   
Eq.\ref{eq:jeans1}-\ref{eq:poisson}  describe the self-consistent Poisson-Jeans equations.
}

\subsection{Dark and baryonic matter distributions in the two models} 

{  
In both the NFW and DMD cases, DM  is not observable, so one must assume its spatial and velocity distributions. In the NFW  case DM is in a  stationary and  spherical configuration, where the gravitational force is in equilibrium with the isotropic velocity pressure. It is important to acknowledge that the idealization of spherical halos and isotropic velocity distributions is a simplification. In reality, halos formed in cosmological simulations exhibit deviations from perfect sphericity, and their velocity distributions are not strictly isotropic. However, for the purposes of the subsequent treatment, it is common to assume a spherical halo and an isotropic velocity distribution. This simplifying assumption allows for a tractable analysis and provides a useful starting point for understanding the overall dynamics of the system.  In  this condition the DM obeys to the Jeans equation in spherical coordinates   \citep{Binney+Tremaine_2008}
}
\be
\frac{1} { \rho_{dm}}  \frac{d \rho \overline{v_{R,\text{dm}}^2} } { dr } +  \frac{ \beta_{\text{dm}} \overline{v_{R,\text{dm}}^2} } { r } = \frac{d \Phi_{\text{dm}} } { dr } 
\ee 
{  
where $r^2=R^2+z^2$, 
and the quantities 
$ \rho_{\text{dm}}$,
$v_{R,\text{dm}}^2$,
$\beta_{\text{dm}}$,
$\Phi_{dm}$
are respectively 
the density, the radial velocity dispersion, the  velocity anisotropy parameter and the  gravitational potential 
of the DM component. The latter quantity is related to the density by the Poisson equation 
}
\be
\label{eq:poisson_nfw_dm}
\nabla^2 \Phi_{\text{dm}} = 4 \pi G \rho_{\text{dm}} \;. 
\ee

{  
Thus in this case Eqs.\ref{eq:jeans1}-\ref{eq:jeans2} become 
}
\bea
\label{eq:jeans1_nfw}
&&
\frac{\partial \rho_{\text{bar}}   \overline{v^2_{R,\text{bar}}} } { \partial R} + 
\frac{ \partial \rho_{\text{bar}}  \overline{v_{R,\text{bar}} v_{z,\text{bar}} }}{ \partial z} 
\\ \nonumber &&
+ \rho \left(
 \frac{
 \overline{v^2_{R,\text{bar}}} 
- \overline{v^2_{\theta,\text{bar}} } } {R} + 
\frac{ \partial \Phi}{ \partial R} \right) =0 \;.
\eea 
{ 
and
}
\bea
\label{eq:jeans2_nfw}
&&
\frac{\partial \Phi} { \partial z} = 
- \frac{\overline{v_{R,\text{bar}}  v_{z,\text{bar}} }} { R} 
- \frac{1} { \rho_{bar} } \frac{\partial \rho\overline{v_{R,\text{bar}}  v_{z,\text{bar}}} } { \partial R} 
\\ \nonumber &&
- \frac{1} { \rho_{bar} } \frac{\partial \rho\overline{v_{z,\text{bar}}^2}} { \partial z} \;,
\eea 
{ 
where $\rho_{\text{bar}} $, $v_{R,\text{bar}}, v_{z,\text{bar}} and v_{\theta,\text{bar}}$ 
are respectively the density and the three components of the velocity  of the baryonic matter in the disk (i.e., stellar and gas components).
Its gravitational potential is 
}
\be
\label{eq:poisson_nfw_bar}
\nabla^2 \Phi_{\text{bar}} = 4 \pi G \rho_{\text{bar}} \;. 
\ee
{  
In order to have a self-consistent Jeans-Poisson system of equations, the total gravitational potential in Eqs.\ref{eq:jeans1}-\ref{eq:jeans2}  is 
}
\be
\label{phi_nfw}
\Phi = \Phi_{\text{bar}} + \Phi_{\text{dm}} \;. 
\ee

{  As with the NFW case, the properties of DM in the DMD model are also unknown and must be assumed. In particular, the hypothesis in the DMD model is that DM is located in the disk and follows approximately the same spatial and velocity distributions as baryonic matter.
Thus in this case  in Eq.\ref{eq:jeans1}-\ref{eq:poisson}  we have that the source of the gravitational potential is the whole matter density that is located in this disk, i.e.,  }
\be
\label{rho_dmd} 
\rho =  \rho_{\text{dm}}+\rho_{\text{bar}} \;, 
\ee
{  and 
$v_R=v_{R,\text{dm}}=v_{R,\text{bar}}$  (the same for other two velocity components).
The total gravitational potential $\Phi$  obtained from the Eq.\ref{eq:poisson} with the density given by Eq.\ref{rho_dmd},   
 coincides with the  gravitational potential of the disk. In this way the  Jeans-Poisson system of equations is again  self-consistent.

In order to use a more compact terminology we define the density in the disk $\rho_\text{disk}$ and the velocity components in the disk, e.g.  $v_{R,\text{disk}}$
where
}   
\bea
&&
\label{NFW} 
\rho_\text{disk} =  \rho_\text{bar}
\\ \nonumber &&
v_{R,\text{disk}} = v_{R,\text{bar}}
\eea 
{ 
for the NFW case and 
}
\bea
&&
\label{DMD} 
\rho_\text{disk} =  \rho_\text{bar}+ \rho_\text{dm} 
\\ \nonumber &&
v_{R,\text{disk}} = v_{R,\text{bar}} = v_{R,\text{dm}}
\eea 
{  
for the DMD case. 
} 

{  By defining the quantities }
\bea
&&
\label{h1} 
\overline{ h_R(R,z)}   = - \left(\frac{1} {\rho_\text{disk}}  \frac{\partial \rho_\text{disk}} { \partial R } \right)^{-1} 
\\ \nonumber && 
\overline{ h_z(R,z)}   = - \left(\frac{1} {\rho_\text{disk}}  \frac{\partial \rho_\text{disk}}  { \partial z } \right)^{-1} \;.
\eea
{ 
 we can rewrite the Jeans equations as  }
\bea
\label{eq:vc_ave}
&&
v_c^2(R,z) = R \frac{\partial \Phi} { \partial R }  \approx \overline{v^2_{\theta,\text{disk}}} - 
\\ \nonumber && 
\overline{v^2_{R,\text{disk}}} \left( 1 -  \frac{R}{\overline{ h_R(R,z)} } -  \frac{R}{h_{v^2_{R,\text{disk}}}} \right) 
\eea
and 
\bea
\label{eq:az_ave}
&&
a_z(R,z)  = \frac{\partial \Phi} { \partial z}  \approx   \frac{\overline{v^2_{z,\text{disk}}}} { \overline{ h_z(R,z)} } - \frac{\partial  \overline{v_{z,\text{disk}}^2} }  { \partial z}   \;, 
\eea
{  Eqs.\ref{eq:vc_ave}-\ref{eq:az_ave} hold for both the NFW and the DMD models, but the disc density is different in the two cases (i.e., Eq.\ref{NFW} and Eq.\ref{DMD} respectively). 

In Eq.\ref{eq:vc_ave}}
where  we have defined 
\be
\frac{1}  {h_{v^2_{R}} } = \frac{1} { \overline{v^2_{R,\text{disk}}}}   \frac{\partial  \overline{v^2_{R}}}{ \partial R} 
 \ee 
 as it is observationally found that (see below)
 \be
\overline{v^2_{R, \text{disk}}}  \propto \exp\left(-\frac{R}{h_{v^2_{R}}}\right)\;.
\ee
In Eq.\ref{eq:vc_ave} we neglect hereafter the mixed terms (i.e., $ \overline{v_Rv_z}$) as they are subdominant. For instance, \cite{Eilers_etal_2019} estimated that the contribution of these terms is 2-3 orders of magnitude smaller compared to other terms. They concluded that these terms introduce systematic uncertainties only at the level of 1\%. In the LIM analysis, estimations of these terms from the data confirm that their contributions are indeed negligible \citep{Lopez-Corredoira_Sylos-Labini_2019, Wang_etal_2023}.

From Eq.\ref{eq:vc_ave} and Eq.\ref{eq:az_ave} it follows that  both the determinations of $v_c$ and $a_z$ from the kinematic data require the knowledge of the density distribution of the disk, as the quantities $\overline{h_R(R,z)}$ and $\overline{h_z(R,z)}$ are involved on the right-hand side of Eq.\ref{eq:vc_ave} and Eq.\ref{eq:az_ave}, respectively. However, the circular velocity is primarily influenced by the tangential velocity $\overline{v^2_\theta}$, and the terms involving $\overline{h_R(R,z)}$ represent only second-order perturbations. On the other hand, the vertical acceleration strongly depends on $\overline{h_z(R,z)}$, as it appears in the denominator of Eq.\ref{eq:az_ave}.

{  Let us now consider a simple case: by approximating the density of the disc as  a double exponential } 
\be
\label{dendobexp} 
\rho_\text{\text{disk}}(R,z) = \rho_0 \exp\left(-\frac{R}{h_R} -\frac{|z|}{h_z}\right) \;,
\ee
 {  
then, from Eqs.\ref{h1}, we simply have 
}
\bea
\label{h2} 
&&
\overline{ h_R(R,z)}   = h_R 
\\ \nonumber && 
\overline{ h_z(R,z)}   =  h_z \;. 
\eea

{  In general,   the density of the disc can be approximated as the } sum of $N_c$ different contributions approximately double-exponentially decaying with different characteristic scales, we have 
\be
\label{multi1}
\rho_\text{\text{disk}}(R,z)
= \sum_i^{N_c} \rho^i_0(R,z) \exp\left(-\frac{R}{h_{R,i}} -\frac{z}{h_{z,i}} \right)
\ee 
where $h_{R,i}$ and $h_{z,i}$ are respectively the characteristic radial and vertical length scales of the $i^{th}$ component.  
{  In this case we have that
\be
\label{multi2}
 \frac{\partial \rho_\text{\text{disk}}} { \partial R }  = -  \sum_i^{N_c} \frac{ \rho^i_0(R,z) \exp\left(-\frac{R}{h_{R,i}} -\frac{z}{h_{z,i}} \right) } { h_{R,i} }  
 \frac{\partial h_{R,i}} { \partial R } 
 \ee
 and 
\be
\label{multi3}
\frac{\partial \rho_\text{\text{disk}}} { \partial z }  = -  \sum_i^{N_c}  \frac{\rho^i_0(R,z) \exp\left(-\frac{R}{h_{R,i}} -\frac{z}{h_{z,i}} \right) } { h_{z,i} } 
\frac{\partial h_{z,i}} { \partial z } \;.
\ee
}
By inserting the values of Eq.\ref{multi1}-\ref{multi3} into Eq.\ref{h1} we can compute the values of   $\overline{ h_R(R,z)} $ and of  $ \overline{ h_R(R,z)} $.

\subsection{Baryonic components} 

{  Let us discuss in more details the properties of the baryonic disc components, that are assumed to be known. These components enter in both  the NFW and DMD mass model. The total gravitational potential of the baryonic components can be written as  
\be
\label{phi_barcomp} 
\Phi_{\text{bar}} = \Phi_{\text{tn}} +  \Phi_{\text{tk}} + \Phi_{\text{bulge}} + \Phi_{\text{\HI}} 
\ee
where 
$\Phi_{\text{tn}}$, $\Phi_{\text{tk}}$,  $ \Phi_{\text{\HI}}$  and  $ \Phi_{\text{bulge}}$  are respectively  the potential of the thin, thick disc and the \HI\; disk 
Similarly the }
total density of the baryonic components is 
\be
\label{rho_barcomp} 
\rho_{\text{bar}} = \rho_{\text{tn}} +  \rho_{\text{tk}} + \rho_{\text{bulge}} + \rho_{\text{\HI}} \;. 
\ee
{  The functional behavior of each of the four baryonic components is given below.}

\subsubsection{Bulge}

The density  of the spherical bulge is typically described by an Hernquist profile \citep{Juric_etal_2008}
 \be
\rho_{\text{bulge}}(r) = \frac{\rho_b^0}{  (r/r_b) \left(1+r/r_b \right)^3 }
\ee
where $r^2=R^2+z^2$ is the 3D radius and $r_b$ the characteristic scale of the bulge. 
The total bulge mass is 
\be
M_{\text{bulge}}=2 \pi \rho_b^0 r_b^3 
\ee
and the gravitational potential is  
\be
\Phi_{{\text{bulge}}} =  -\frac{GM_{b}}{r_{b}} \frac{1}{1+\frac{r}{r_b}} \;. 
\ee
The characteristic length scale is $R_b = 0.25$ kpc and its mass is $M_{\text{bulge}} = 2 \times 10^{10} M_\odot$  \citep{Juric_etal_2008}.

\subsubsection{Thin disc} 

The density of the thin disk can be approximated by a double-exponential 
\be
\label{rho_thin} 
\rho_{\text{tn}}(R,z) = \rho_{tn,0}   \left( -\frac{R}{h_{tn,R}} \right)
\exp\left(-\frac{|z|}{2h_{tn,z}(R)}\right)
\ee
where $\rho_{tn,0}   $ is a constant and  the vertical thickness $h_{tn,z}(R)$ depends on the radius, i.e., there is a flare whose radial behavior can be fitted as \citep{Chrobakova_etal_2022} 
\be
\label{hz_thin} 
h_{tn,z}(R)  = 0.14 - 0.0037 R + 0.0017 R^2\;.
\ee
The radial characteristic length scale of the thin disc is $h_{tn,R} = 4.5$ kpc and mass $M_{\text{tn}} = 3 \times 10^{10} M_\odot$: these are the same parameters used by \cite{Eilers_etal_2019,SylosLabini_etal_2023_MW}.

\subsubsection{Thick disc} 

Even for the thick disk the density can be approximated by a double-exponential
\be
\label{rho_thick} 
\rho_{\text{tk}}(R,z) = \rho_{tk,0} \left( -\frac{R}{h_{tk,R}} \right) 
\exp \left(- \frac{|z|}{2h_{tk,z}(R)}\right)
\ee
where $\rho_{tk,0}   $ is a constant  and the vertical thickness $h_{tn,z}(R)$ depends on the radius. The flare in this case is described as \citep{Chrobakova_etal_2022}
\be
\label{hz_thick} 
h_{tk,z}(R)  = 1.21 - 0.19  R + 0.015 R^2 \;. 
\ee
The radial characteristic length scale of the thick disc is $h_{tk,R} =2.3 $ kpc   and mass $M_{\text{tk}} = 2.7 \times 10^{10}  M_\odot$,
where even in this case we have adopted  the same parameters as \cite{Eilers_etal_2019,SylosLabini_etal_2023_MW}.

\subsubsection{The neutral hydrogen disk} 

We assume the density of the gaseous disk as the one 
derived for \HI, that is the dominant gas component.  \cite{Kalberla_Dedes_2008} found that this can be approximated as 
 \be
\label{rhogas} 
\rho_{\text{\HI}} (R,z)  =  \rho _{g,0} \exp \left( -\frac{R}{h_{g,R} } - \frac{|z|}{2h_{g,z}(R)}\right)
\ee
where $\rho _{g,0} $ is a constant  and 
\begin{equation}
\label{hzgas}
h_{g,z}(R) =(0.15\ {\rm kpc})\exp \left( \frac{R-R_\odot}{h_{R,fg}}\right)
\end{equation}
and where  $h_{R,fg}=9.8$ kpc. The relative error on $h_{g,z}$  is about 20\% \citep{Kalberla_Dedes_2008}.  The total \HI \; mass is $M_g=0.5 \times 10^{10} M_\odot$.  The flares of the gas and thin disks are similar and smaller than that of the thick disk.

\subsection{Dark matter in the NFW model}

The DM halo is assumed to have the canonical NFW density profile that is defined by two parameters, a characteristic length $r_s$  and amplitude $\rho_{\text{h}}^0$ 
\be
\label{rho_nfw} 
\rho_{\text{dm,halo}}(r) = \frac{\rho_{\text{h}}^0} {\left(1+\frac{r}{r_s} \right)^2 \frac{r}{r_s}} \;.
\ee
Sometimes different shapes are used: for instance, \cite{Ou_etal_2024} considered a generalized NFW and an Einasto profile which have an additional parameter. For simplicity we will consider Eq.\ref{rho_nfw} only as even the DMD model is characterized by two free parameters that must be constrained from the best fitting the data.

The two free parameters of the NFW can be expressed in terms of the virial radius and mass \citep{Navarro_etal_1997}. 
The virial radius $r_{\text{vir}}$ is defined as the radius  at which the average density within this radius is $\Delta =200$  times the critical or mean
density of the universe.  By defining  $c=r_{\text{vir}}/r_{s}$ as the  concentration parameter the virial mass,$M(r_{\text{vir}})$, i.e., the mass inside the virial radius $r_{\text{vir}}$, is 
\be
M(r_{\text{vir}}) = 4 \pi \rho_{\text{h}}^0 r_s^3 \left( \log(1+c) - \frac{c}{1+c} \right) \;.
\ee
Given the spherical symmetric nature of the halo gravitational potential can be analytically  calculated \citep{Navarro_etal_1997}.

\subsection{Dark matter in the DMD model} 

As previously mentioned, in the analysis of external galaxies, it is assumed that the DM profile is a rescaled version of those derived from the \HI{} distribution \citep{Sancisi_1999, Hoekstra_etal_2001}.
The rationale behind this choice is as follows: in the context of external galaxies, it is observed that the surface density of \HI{} decays at a slower rate compared to that of the stellar component. Consequently, the rotation curve attributed to the gas alone exhibits a much slower decay than that of the stellar component. Typically, the characteristic length scale for the exponential decay of the gaseous component can be around five times larger than that of the stellar component.
Therefore, when appropriately rescaled, the rotation curve of the gas allows for a range of observed rotation curve shapes. This includes nearly flat rotation curves, as well as cases where the rotation curves decay similar to that of the MW, or even increase with radial distance. The additional matter that needs to be postulated to align with the observed rotation curves, especially at sufficiently large radii, is approximately tens of times greater than that of the neutral \HI{}.

In addition, a part of DM can be also associated with the stellar component. Indeed, it has been shown that the Bosma effect can provide highly accurate rotation curve fits for several disc galaxies, by using both the observed stellar disc and \HI{} gas as proxies, with different weights, for DM \citep{Hessman+Ziebart_2011, Swaters_etal_2012,SylosLabini_etal_2024_Mass}. Thus, in general the DMD fits uses two free parameters  corresponding to the weights associated to the stellar and gas components. 

{  Here, we consider the radial length scale, $\overline{h_{R}}$, and the dark matter mass, $M_\text{dm,disk}$, as free parameters of the DMD model. This approach assumes that the properties of the stellar components, which have been accurately measured for the MW, are fixed, and that DM is associated only with the neutral hydrogen component. In this case, the parameter $\overline{h_{R}}$ is expected to be of the same order as the \HI{} length scale. However, it is treated as a free parameter because the observed \HI{} distribution exhibits a more complex behavior than a simple exponential function \citep{Kalberla_Dedes_2008,SylosLabini_etal_2023_MW}. These two parameters can be determined by fitting the rotation curve on the plane. The vertical density distribution needs to be computed numerically once the values of $\overline{h_{R}}$ and $M_\text{dm,disk}$ have been fixed.} 

\section{Results} 
\label{gaiadr3} 

We begin by discussing the estimation of the kinematic moments using data from Gaia-DR3. In particular, we briefly describe the methodology and techniques employed to extract the necessary information from the Gaia data, including the selection criteria for the sample, data processing, and measurement techniques for obtaining the kinematic moments.
We then present the on-plane fits achieved with the two mass models that we have chosen. 
{  By using the Jeans equations, we compare the derivative of the gravitational potential derived by these models with the observed kinematic moments.}

\subsection{The Lucy's inversion method} 

{  The Gaia-DR3 stellar sample is limited to $R=14$ kpc in radial distance and consists of $\sim 1.6$ millions of stars with the 6D coordinates. Indeed, for $R>14$ kpc  the the relative error in the
distance becomes larger  than 20\% \citep{Gaia_2021}.  This sample, with a cut in the vertical height to $|z|<2 $ kpc, is useful to study the Galactic region close to the Sun. To expler larger radial distances we will use kinematic maps reconstructed by 
\cite{Lopez-Corredoira_Sylos-Labini_2019}. In particular, they have obtained 
}  
 extended kinematic maps of the Galaxy using Gaia DR2 data, specifically targeting the region where the relative error in distance ranged from 20\% to 100\%. To achieve this, they employed Lucy's inversion method (LIM) \citep{Lucy_1977}, developing a statistical deconvolution algorithm for parallax errors. By applying LIM to the Gaia DR2 dataset and incorporating line-of-sight velocity measurements, they extended the distance range for kinematic analyses by approximately 7 kpc compared to the results presented by \cite{Gaia_2018}. This extension included Galactocentric distances ranging from 13 to 20 kpc, providing valuable insights into the kinematics of the MW in those regions.   \cite{Wang_etal_2023}, who applied the same LIM to the Gaia DR3 sources reaching radial distances of 30 kpc. Their findings were consistent with the results obtained from applying LIM to Gaia DR2 sources, confirming that LIM yields convergent and more accurate results by improving the dataset's statistics and reducing observational errors. The kinematic maps reconstructed using LIM, covering distances up to approximately 30 kpc, revealed asymmetrical motions with significant velocity gradients in all components. These observations highlight the complex and dynamic nature of the MW.

The LIM provides estimates of the  velocity components, along with their corresponding errors and rms values, for a certain number of cells ($N_{\text{cells}}$) into which the Galactic region is divided. The deconvolution process includes all stars that have parallax errors smaller than the parallax itself and are located within a {  galactic latitude range} of $|b| < 10^\circ$, i.e. the anti-center region. To ensure a sufficient number of stars for reliable estimates, the Galactic region meeting the above criteria is further divided into 36 line-of-sight cells. Each of these cells has a size of $\Delta \ell = 10^\circ$ in galactic longitude. The deconvolution technique discussed earlier is then applied to each of these cells. It is important to note that only cells with a number of stars greater than six (i.e., $N > 6$) are considered in the subsequent analysis. This criterion ensures that there is a minimum number of stars available in each cell to obtain meaningful and statistically robust results.

 \cite{Wang_etal_2023}  {  divided} the anti-center region in 24448 cells of size $\Delta R=\Delta z=0.15$ kpc: with respect to that sample  we have eliminated  high velocity stars imposing the following limits to the velocity components $v_\theta \in(0,400)$ km sec$^{-1}$,  $|v_R|<100$ km sec$^{-1}$  and $|v_z|<100$ km sec$^{-1}$: for this reason we have less cells, i.e., 14101,  which cover a smaller range of radial and vertical distances. {  In addition, we} require that  -2 kpc $ \le z \le $ 2 kpc and {  that} 8.5 kpc $\le R \le$ 25 kpc: {  with} these constraints we are left with 3201 cells.  In \cite{Wang_etal_2023} the rotation curve was determined  in the range of vertical heights  {  up to 2 kpc}: here we reduce the the radial range of distances in order to have a more robust statistics in the 4 slices with different vertical heights that we are going to consider, i.e.  $|z| \in [0,0.5], [0.5;1], [1,1.5]$ and $[1.5,2]$,  kpc. In what follows, each vertical slice is identified by the mean {  vertical height}, i.e. $|z|=0.25, 0.75, 1.25. 1.75$ kpc.

\subsection{Estimation of kinematic moments from Gaia-DR3} 
\label{estDR3}

The average values of the three velocity components are consistent with the determinations of \cite{Wang_etal_2023}: their behavior aligns with other results available in the literature, and we refer the interested reader to that work for further details.
{  
As an additional test, let us consider the comparison between the direct measurements in the Gaia-DR3 sample and the mesurements derived from the LIM-reconstructed data in the same region of the Galaxy. Fig.\ref{Vphi_Gaia+LIM} shows the case of the tangential velocity $v_\phi(R,z)$ and Fig. \ref{sigmaVr} presents the behavior of the radial velocity dispersion $\sigma_{v_r}(R,z)$: both have been computed as a function of the radial distance  in bins $\Delta z=0.5$ kpc. 
Fig.\ref{sigmaVz}   shows the vertical velocity dispersion $\sigma_{v_z}(R,z)$ vs $|z|$  in bins $\Delta R=2$ kpc.

 In the case of the Gaia-DR3 sample  the error bars on the radial and vertical velocity dispersions have been computed using bootstrap resampling whereas for the case of the LIM results by propagating the errors given by the reconstruction method \cite{Wang_etal_2023}. 
  It is noteworthy that these measurements are consistent with each other within the error bars  for $v_\phi(R,z)$ and $\sigma_{v_z}(R,z)$, whereas the difference between the two determinations of $\sigma_{v_r}(R,z)$ in the first   distance bins is due to the different limits on the maximum value of the radial velocity used in the two samples (i.e., no limits vs $|v_r|<100$ km s$^{-1}$).   

We observe that $v_\phi(R;z)$ shows a clear transition from a decreasing to an increasing trend in the range $R<15$ kpc, moving from measurements on the galactic plane (i.e., $z<0.25$ kpc) to regions above it (i.e., $z=1.75$ kpc). 
{  Furthermore, we note that the results for $\sigma_{v_r} \approx \sqrt{\overline{v^2_{R,\text{disk}}}}$ 
approximately coincide 
with the findings of \cite{Eilers_etal_2019, Ou_etal_2024} on the galactic plane, who have fitted the behavior using with the  exponential function
}
\be
\sqrt{\overline{v^2_{R,\text{disk}}}} \propto \exp \left( - \frac{R}{25} \right) \:. 
\ee

\begin{figure} 
\includegraphics[width=0.5\textwidth]{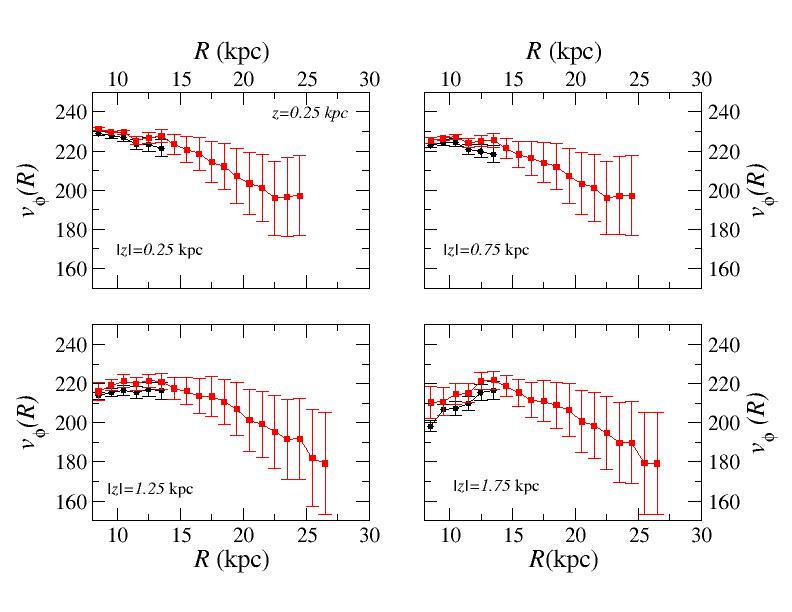}
\caption{  
Behavior of $v_{\phi} (R,z)$ vs $R$  in bins $\Delta z=0.5$ kpc. Black circles are the direct measurements in the Gaia DR3 sample
(that are limited to $R<12$ kpc), while red squares are the determinations through the LIM reconstructed data in the same region of the Galaxy.  
} 
\label{Vphi_Gaia+LIM} 
\end{figure}

\begin{figure} 
\includegraphics[width=0.5\textwidth]{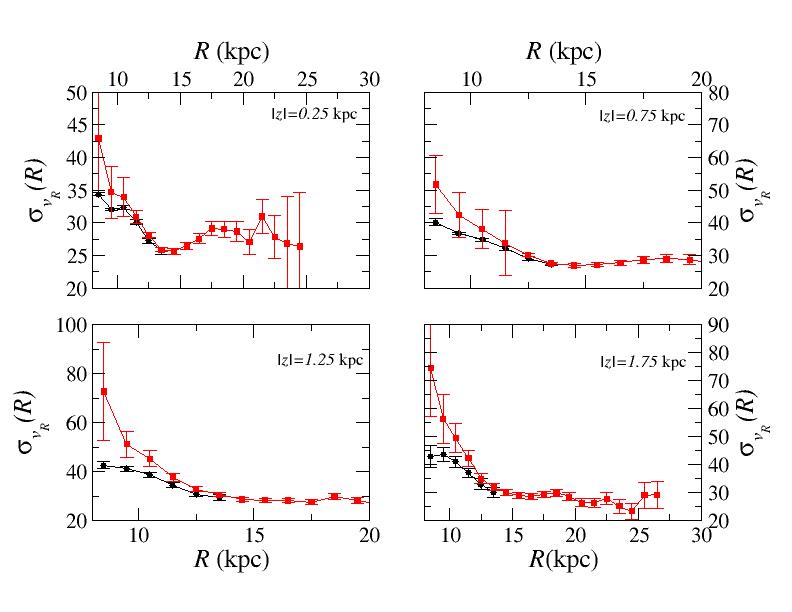}
\caption{  
Behavior of $\sigma_{v_r}(R,z)$ vs $R$  in bins $\Delta z=0.5$ kpc. Black circles are the direct measurements in the Gaia DR3 sample {  (that are limited to $R<12$ kpc),} while red squares are the determinations through the LIM reconstructed data in the same region of the Galaxy. 
} 
\label{sigmaVr} 
\end{figure}

Even determinations  of the vertical velocity dispersion $\sigma_{v_z} (R,z)$  vs $z$ in radial  bins of width $\Delta R=2$ kpc show reasonable agreement, in the range of radii where they overlap, of the direct measurements in the Gaia-DR3 catalog and in the LIM analysis. 
\begin{figure} 
\includegraphics[width=0.5\textwidth]{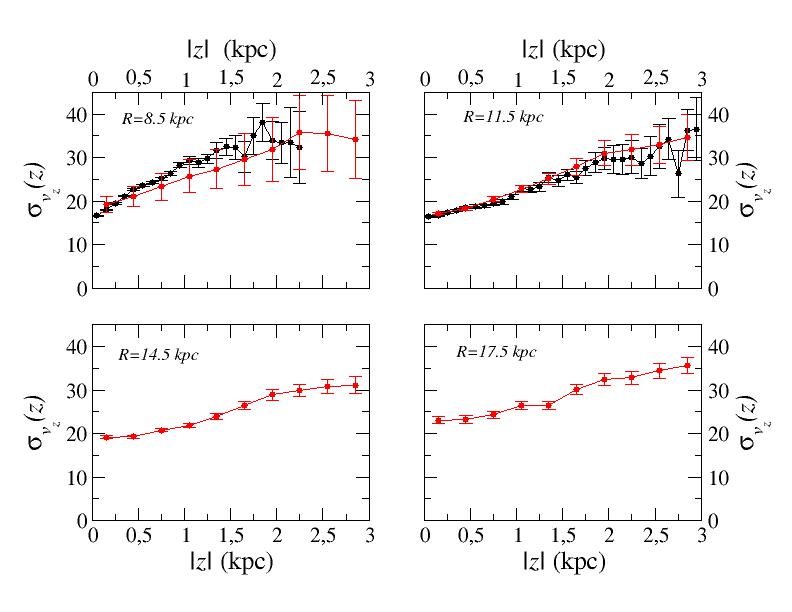}
\caption{ 
Behavior of $\sigma_{v_z}(R,z)$ vs $|z|$  in bins $\Delta R=2$ kpc.  Black circles are the direct measurements in the Gaia DR3 sample
(that are limited to $R<12$ kpc), while red squares are the determinations through the LIM reconstructed data in the same region of the Galaxy. 
 } 
\label{sigmaVz} 
\end{figure}

Fig.\ref{sigmaVz2} shows the behavior of $\sigma_{v_z}  (R,z)$ vs $z$  in bins $\Delta R=3$ kpc centered in  
$R=8.5, 11.5, 14.5, 17.5, 20.5$ kpc 
and $\Delta z=0.3$ kpc. One may note that  $\sigma_{v_z} (R,z)$ approximately growths linearly with $z$ and that its amplitude a low $z$ slightly increases with $R$. 
\begin{figure} 
\includegraphics[width=0.5\textwidth]{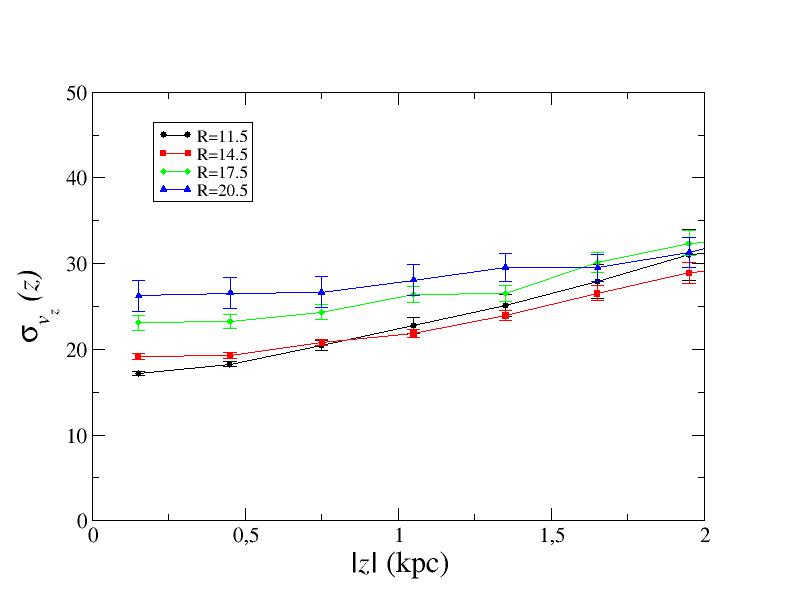}
\caption{  
Behavior of $\sigma_{v_z} (R,z) $ vs $|z|$  in bins $\Delta R=3$ kpc centered in  
$R=8.5, 11.5, 14.5, 17.5, 20.5$ kpc 
and $\Delta z=0.3$ kpc. } 
\label{sigmaVz2} 
\end{figure} 
We have fitted to the estimated values of $\sigma_{v_z}  $ vs $z$ the function 
\be
\label{sigmavz_fit_3} 
\sigma_{v_z} (R;z)  =  C_0(R) + C_2(R) z^2 \;,
\ee
where the functional
behavior is chosen so that the first derivative of $\sigma_{v_z} (R;z)$ for $z=0$ is zero because of symmetry reasons. 
\begin{table}
\begin{center}
\begin{tabular}{ | c | c | c |   }
\hline 
 $R$	&	$C_0$&	$C_2$ \\ 
\hline
 11.5&	$19.7 \pm 2.0$ &	$2.1 \pm 0.2$\\
 14.5&	$20.4 \pm 1.2$ &	$1.6 \pm 0.1$ \\
 17.5&	$24.3 \pm 1.2$ &	$1.5 \pm 0.1$ \\
 20.5&	$26.5 \pm 1.4$ &	$1.5 \pm 0.1$ \\
\hline
\end{tabular}
\end{center}
\caption{
The values of the coefficients, along with their respective errors, for the second-order polynomial used to fit $\sigma_{v_z} (R;z)$ as a function of  $z$ (see Eq.\ref{sigmavz_fit_3}).
}
\label{tab_coeff_3}
\end{table}
{  The values of the coefficients and their standard errors in both cases are reported in Tab.\ref{tab_coeff_3}. }

Fig.\ref{Coeff} shows the behavior of $C_0(R)$ and $C_2(R)$ of Eq.\ref{sigmavz_fit_3} with their linear fits 
\bea
\label{eq:coeff_3} 
&&
C_0(R) = (9.8 \pm 2.0)  + (0.80 \pm 0.14) R
\\ \nonumber &&
C_2(R) = (2.6 \pm 0.4)  - (0.86 \pm 0.06)  R \;. 
\eea
\begin{figure} 
\includegraphics[width=0.5\textwidth]{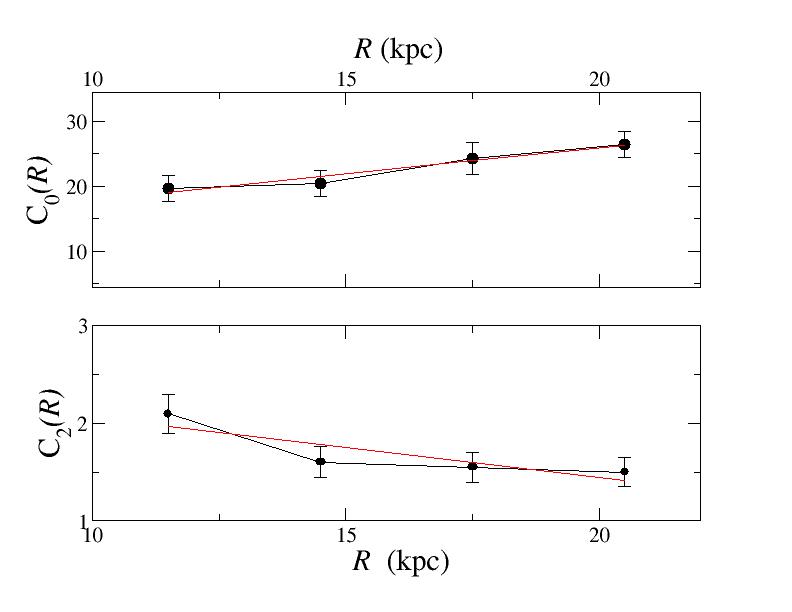}
\caption{  Behavior of the two  coefficients describing the behavior  $\sigma_{v_z}(z;R)$ in function of $|z|$ (see Eq.\ref{sigmavz_fit_3}). Best fits with a linear function (red lines) are reported (see Eq.\ref{eq:coeff_3}).} 
\label{Coeff} 
\end{figure} 
}

\subsection{Numerical determination of the gravitational potential} 

{  The best fit to either the DMD or NFW model on the galactic plane was obtained by adjusting two free parameters.
As mentioned above, for the DMD model, these parameters are the characteristic length scale of the radial exponential decay, $\overline{h_{R}}$, and the mass of the DM component $M_{\text{dm,disk}}$. By varying their values, the DMD model can be adjusted to achieve the best fit to the observed data on the galactic plane. On the other hand, for the NFW model the two free parameters are  the characteristic length  $r_s$, and the amplitude $\rho_h^0$ (see Eq. \ref{rho_nfw}). 
While the expression of halo potential can be analytically computed for any value of the spherical distance, the contribution of the disk components can be easily computed only on the plane. 
}

When it comes to fitting the off-plane behavior of  the generalized rotation curves $v_c(R,z)$ and the vertical acceleration $a_z(R,z)$, the task becomes more challenging due to the lack of analytical expressions for these quantities in the 
{  baryonic} or DMD disks. As a result, two different approaches can be employed:
i) One approach is to compute the derivative of the gravitational potential through numerical integration, as implemented by \cite{Zhu_etal_2023}. 
ii) Another approach is to perform numerical realizations of the DMD disk or halo model and calculate the potential in such particle distributions. In this study, the second approach was adopted, although it does not allow for a complete best-fitting procedure. Instead, multiple realizations of the DMD disk or halo model were run by varying the two free parameters to find the best fit to the data.

 {  In each realization particles are spatially distributed according to the  density distributions corresponding to the two models. More specifically, the density of  the baryonic disc components is described by Eq.\ref{rho_barcomp} while the DM component is given by Eq.\ref{rho_nfw} for the NFW halo case or given by Eq.\ref{rhogas} for the DMD case. Each realization of the whole baryonic+DM system consists of  $N_p \approx 5 \times10^6$ particles of equal mass $m$.  The  gravitational potential (for unit mass) of the $i^{th}$ particle, in the position  $\vec{x}_i$ is computed as }
\be
\Phi_i (\vec{x}_i) = \sum_{j=1;  j\ne i }^{N_p} \frac{G m}{ |\vec{x}_i-\vec{x}_j|} \;, 
\ee
{  where $G$ is the Newton's constant. The mean gravitational potential is then calculated as }
\be
\label{Phi_num_R}
\overline{\Phi(R;\Delta R; z_1<z<z_2) } = \frac{1}{N_s} \sum_{i=1}^{N_s} \Phi_i (\vec{x}_i) 
\ee
{  
where the sum is extended to the $N_s$ in the bin of size $\Delta R$ centered at $R$ and with $z \in [z_1, z_2]$. 
The radial derivative of Eq.\ref{Phi_num_R} gives an estimation of the total gravitational potential of a given mass model 
that enters in  Eq.\ref{eq:vc_ave} (the same reasoning applies for the vertical derivative that enters in Eq.\ref{eq:az_ave}). 
The kinematic moments in both Jeans equations are computed in the same $R,z$ bin of the gravitational potential. 
 }

\subsection{Effect of the flare on the radial and vertical accelerations} 

{ 
To numerically determine the gravitational potential and its radial and vertical derivatives for computing the expected  $v_c(R,z)$ and  $a_z(R,z)$, we initially considered, as an illustrative example, a simplified disk with a total density described by a double exponential decay in the radial and vertical density distributions described by Eq.\ref{dendobexp}, instead of the full baryonic density  components (see Eq.\ref{rho_barcomp}).  This example is useful to single out the effect of the flare.
} 

The parameters used, {  which are not chosen to fit  the Gaia-DR3 data,}   are  $M_{\text{disk}}=10 \times 10^{10} M_\odot$, $h_{R}=5$ kpc, and $h_{z}=0.3$ kpc. Additionally, we explored a scenario where the flare was determined using the expression of the flare for the thick disk case (Eq. \ref{hz_thick}), resulting in the largest flare among that of the thin, thick and gas disks.

\begin{figure} 
\includegraphics[width=0.5\textwidth]{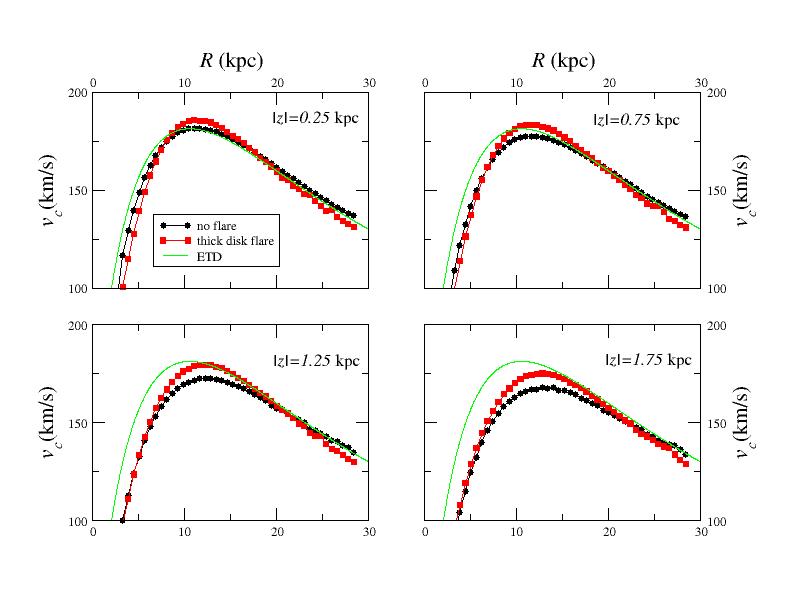}
\caption{  
Behavior of $v_c(R,z)$ for a disk of mass $M = 10 \times 10^{10}  M_\odot$  with a double exponential decay with $h_{g,R}=5$ kpc and $h_{g,z}=0.3$ kpc  (black circles) and for the case where there is a thick-disk-like flare (red circles). The plot is in terms of the radial distance $R$ and is divided into vertical slices of thickness $\Delta z=0.5$ kpc centered at $|z|=0.25$ kpc, ..., $z=1.75$ kpc.  Additionally, the behavior of the exponential thin disk (ETD) model with the parameters of the disk is shown as a reference (represented by a green line). Error bars are reported but not visibile in the figure.
}
\label{gas1a} 
\end{figure}

Figure \ref{gas1a} illustrates the behavior of $v_c(R,z)$ in vertical slices with a thickness of $\Delta z=0.5$ kpc, centered at $z=0.25, 1.25, 2.25$ kpc. In {  all} cases, the behavior of $v_c(R,z=0.25 \; \text{kpc})$ closely resembles that of the exponential thin disk (ETD) model (see \cite{Binney+Tremaine_2008}  Eq.(2.165)), with parameters $M_{\text{disk}}=10 \times 10^{10}  M_\odot$ and $h_{R}=5$ kpc. However, the presence of the flare becomes particularly noticeable around the maximum value of the rotational velocity, that occurs for $R_{\text{max}} \approx 2 h_{R} \approx 10$ kpc. 

This difference is the reason behind the distinct best-fit parameters for the DMD model obtained in the present work (see below) compared to those obtained by \cite{SylosLabini_etal_2023_MW}.
It is worth mentioning that as $z$ increases beyond 0.25 kpc, the behavior of $v_c(R,z)$ deviates for $R<R_{\text{max}}$, exhibiting a faster decay at smaller radii. However, for $R>R_{\text{max}}$, the behavior converges to the ETD model.

\begin{figure} 
\includegraphics[width=0.5\textwidth]{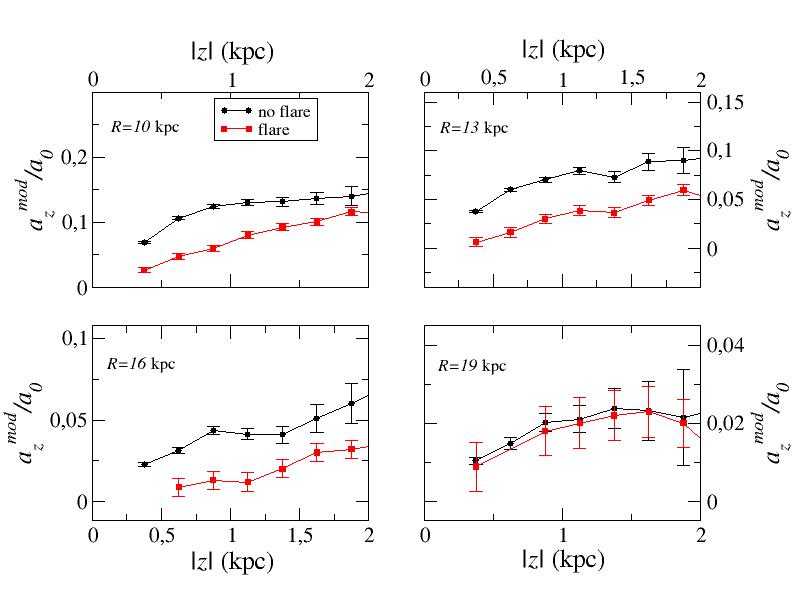}
\caption{  Behavior of $a_z^{\text{mod}} (R,z)$ in units of $a_0$ (see Eq. \ref{a0}). The plot is in terms of the vertical height $|z|$ and is divided into radial slices centered at $R=10$ kpc, $R=13$ kpc, $R=16$ kpc, and $R=19$ kpc. The thickness of each radial slice is $\Delta R=1$ kpc.
}
\label{gas1b} 
\end{figure}

Figure \ref{gas1b} presents the behavior of the vertical acceleration, obtained as 
\be
a_z^{\text{mod}}  (R,z) = \frac{\partial \Phi(R,z)}{\partial z}
\ee  
in function of  $z$ in different radial slices of thickness $\Delta R=$ 1 kpc. Note that $a_z^{\text{mod}} $ has been normalized to
\be
\label{a0} 
a_0 = \frac{GM_{\text{disk}}}{h_{R}^2} \;.
\ee
The amplitude of $a_z^{\text{mod}} $ decreases as the radial distance $R$ increases. Specifically, it follows a decay proportional to $R^{-2}$, which is in line with the expected behavior.
Furthermore, it presents a slow growth with $z$,  and it shows a factor not larger than 1.5 difference between the cases with and without the flare.
In brief, the effect of the flare is that of deforming the behaviors of $v_c(R,z)$ and $a_z(R,z)$ w.r.t. the case a simple double exponential decay. 
{  The vertical acceleration is primarily determined by the disk alone as the contributions from the spherical halo are negligible for distances $|z| < 2$ kpc, due to fact that the halo is spherically symmetric}. 

In the case of the DMD scenario, the mass of the disk is approximately twice that of the NFW scenario. {  This is because, in the NFW case, only the baryonic components contribute to the disk's mass, whereas in the DMD case, there is an additional contribution from DM, which is approximately twice the mass of the baryonic component (as described below).} Since the vertical acceleration ($a_z$) scales with the mass of the disk, we expect the acceleration to be approximately double in the DMD scenario compared to the NFW scenario.

{  We stress that Figs. \ref{gas1a}-\ref{gas1b} present a simple example aimed at illustrating the trends when varying the range of vertical and radial distances for the generalized rotation curve and vertical acceleration, respectively. As mentioned above, since the main contribution to these trends in both the DMD and NFW models comes from the disk, we expect similar behaviors in these models. The only difference lies in the amplitude of  $v_c(R,z)$ and $a_z(R,z)$, which is determined by the mass of the disc.
}

\subsection{A simple exponential disk model} 

\begin{figure} 
\includegraphics[width=0.5\textwidth]{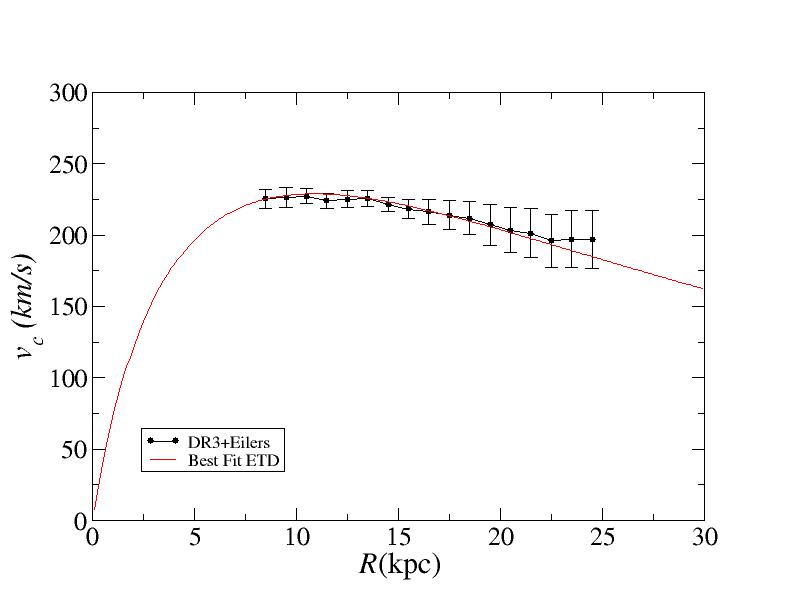}
\caption{
Best fit to the Gaia-DR3 data  in the range of radial distances $[12,22]$ kpc and of vertical distances $|z| \le 0.25$ kpc  with the exponential thin disk approximation (ETD). Best fit parameters are  $M_{\text{disk}} = 17.1 \times 10^{10} M_\odot$ and $h_R = 5.1$ kpc}
\label{Fit_ETD} 
\end{figure}

{   
\cite{SylosLabini_etal_2023_MW} considered the fits to the rotation curve on the plane with two different models: both have the same baryonic components that are described  by Eq.\ref{rho_barcomp} 
with total baryonic mass $M_{\text{bar,disk}} = 8.2 \times 10^{10} M_\odot$. Additionally, in the NFW model the  halo component has  the best-fit parameters i.e. $r_s=12.6 $ kpc and $\rho^0_{\text{h}}=9.4 \times 10^{-25}$ gr cm$^{-3}$ which correspond to a virial mass of $M_{\text{vir}} = 6.5 \times 10^{11} M_\odot$ whereas for the DMD   the additional DM  mass is $M_{\text{dm, disk}} = 8.9 \times 10^{10} M_\odot$.

Fig. \ref{Fit_ETD} shows, within the range of radial distances $[8.5,25]$ kpc and vertical distances $|z| \le 0.25$ kpc, a fit using a much-simplified disk model. This model consists of an exponential thin disk characterized by two parameters: the mass and the radial length scale, with the best-fit parameters found to be  $M_{\text{disk}} = 17.1 \times 10^{10} M_\odot$ and $h_R = 5.1$ kpc respectively. 

These parameters slightly differ from the best fit obtained by \cite{SylosLabini_etal_2023_MW} for several reasons. Firstly, in their study, the disk was modeled using the full stellar components, whereas in our case, we employ a different modeling approach consisting in the ETD approximation. Additionally, the range of radial distances considered in their study was different, spanning from 5 kpc to 27.5 kpc. In contrast, we limit our analysis to the range of 8.5 kpc to 25 kpc. This narrower range is chosen because it allows us to compute the generalized rotation curves at different vertical heights in the same range of radial distances.}

{  Concerning the vertical acceleration, this is determined by two parameters: the mass of the disk and its vertical characteristic length scale, $\overline{h_z}$. The disk mass in the DMD case is about twice that of the NFW model and is fixed by fitting the on-plane rotation curve. Thus, for $a_z(R,z)$ in the example considered, the only free parameter is  $\overline{h_z}$. We find that for $\overline{h_z} = 0.1$ kpc  tthe DMD model agrees with the data, while the NFW model does not. It is worth noting that if we had considered $\overline{h_z} = 1$ kpc, the situation would be reversed. Indeed, as discussed in the next section (see Fig.\ref{az_DR3+Gas1_c}), the amplitude of the vertical acceleration decreases with an increase in $\overline{h_z(R)}$.}

This discussion highlights the importance of the vertical height as a key parameter in assessing the agreement between a model and the vertical acceleration data. Additionally, the presence of different flares in the various components of the disk further complicates the problem, as we will discuss in the following paragraphs. Therefore, it is necessary to construct a realistic model in which all parameters are carefully constrained based on the available data.
At small radial distances, it is required to have a very careful characterization of the flares in the different mass components. However, at large radial distances, i.e., $R>15$ kpc, the generalized rotation curves are weakly dependent on the details of the flares. Thus, this is the more robust radial range to fit theoretical mass models.


\subsection{Determining the observed vertical acceleration} 
\label{sect:az}

\begin{figure} 
\includegraphics[width=0.5\textwidth]{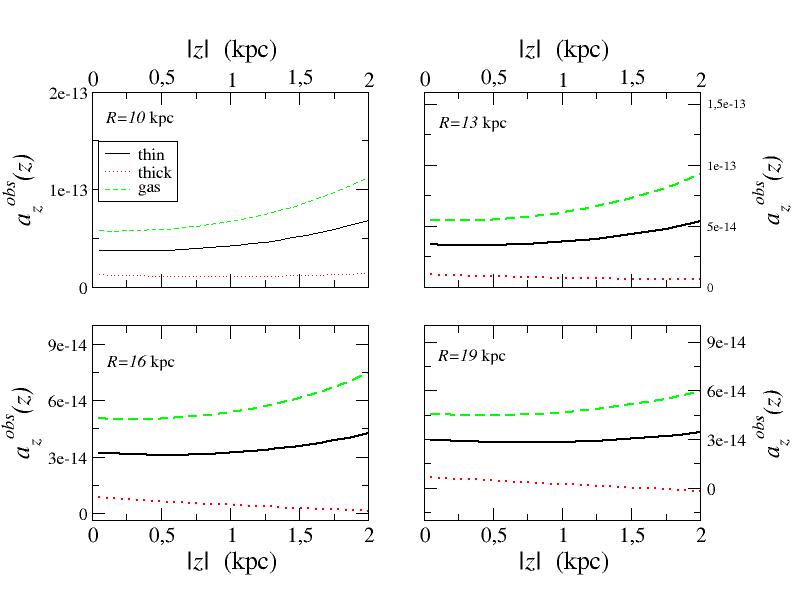}
\caption{Behavior of  $a_z^{\text{obs}}$ (see Eq.\ref{eq:az_ave}) in units of $a_0$ (see Eq.\ref{a0})
for the three cases $\overline{h_z(R)}=h_{tn,z}$, $\overline{h_z(R)}=h_{tk,z}$ and $\overline{h_z(R)}=h_{g,z}$. For clarity the  error bars are not reported.  }
\label{az_DR3+Gas1_c} 
\end{figure}

{  As mentioned above, the two free parameters of the DMD model,  $h_{g,R}$ and $M_{\text{dm,disk}}$,  and of the NFW model, $r_s$ and $\rho_0$, can be  determined from the fit of $v_c(R,z)$ on the plane. 
In the first Jeans equation, that gives $v_c(R,z) $, the model-dependent term that includes the radial characteristic length scale can be treated as a perturbation.
On the other hand, in the second Jeans equation (Eq.\ref{eq:az_ave}) providing $a_z(R,z) $ the term $\overline{h_z(R)}$ becomes crucial and needs to be determined for a specific mass model. 

For the NFW model, $\overline{h_z(R)}$ does not depend on the parameters of the fit but only on the assumed mass of the baryonic components and on their vertical distributions that are all known.   
 In the case of the DMD model,  $\overline{h_z(R)}$  depends on the total mass of the disk, which includes both baryonic and dark matter, as well as on its vertical density distribution, specifically influenced by the characteristics of both the baryonic and DM flares. Given that the baryonic properties are known, ideally, the shape of the DM flare should be determined through a best-fit procedure. However, in this study, a complete best-fitting procedure is not employed. Fig.\ref{az_DR3+Gas1_c} shows the variation in $a_z^{\text{obs}}$ hen fixing the mass of the disk and allowing $\overline{h_z(R)}$ to vary. It can be observed that transitioning from $\overline{h_z(R)}=h_{g,z}$ to $\overline{h_z(R)}=h_{tk,z}$  causes $a_z^{\text{obs}}$ vo change by a factor 10. 
}


\subsection{The DMD model}

\begin{figure}
\includegraphics[width=0.5\textwidth]{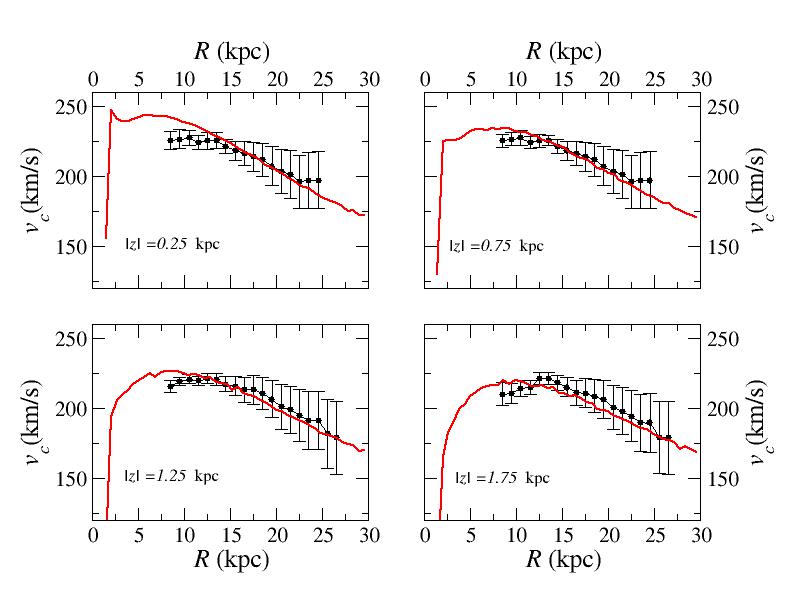}
\caption{  Generalized rotation curves of the Gaia DR3 sample  for different values of the median height (black dots)  and  the best fits with a  DMD model (red lines).
 The DM disk has parameters $M_{\text{dm,disk}} = 9 \times 10^{10}  M_\odot$ and $h_{g,R} = 5$ kpc .
    \label{DR3_Bosma}}
\end{figure}

The best fit results to the generalized rotation curves $v_c(R,z)$ of the Gaia DR3 sample  obtained using a DMD disk model are presented in Fig. \ref{DR3_Bosma} in the range [8.5,25] kpc. The DM disk has  $M_{\text{dm,disk}} = 9 \times 10^{10} M_\odot$ and $h_{g,R} = 5$ kpc. Note that the baryonic components have mass $M_{\text{bar}} = 8.2 \times 10^{10} M_\odot$ so that the total mass is very similar to the single disk $M_{\text{disk}}=17.2 \times 10^{10}  M_\odot$  of the ETD fit presented in the previous section as it is similar $h_{g,R} $. A safe estimate of the uncertainty in the mass estimation is about 10\%.

\begin{figure}
	\includegraphics[width=0.5\textwidth]{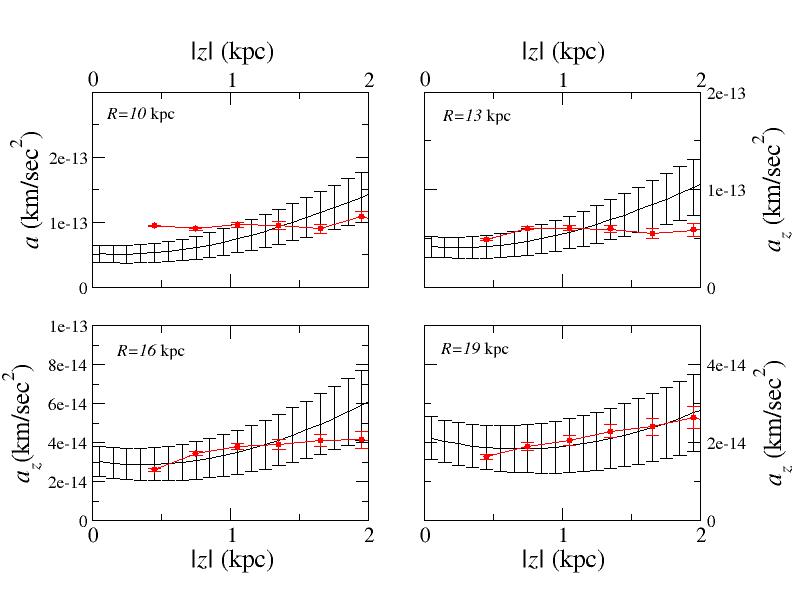}
	\includegraphics[width=0.5\textwidth]{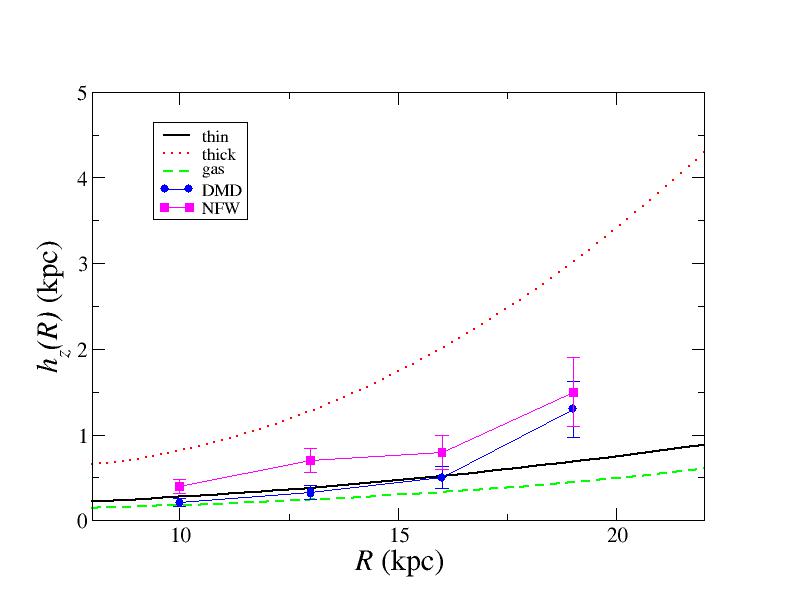}
 \caption{  Upper panel: behaviors of $a_z^{\text{mod}}(R,z)$ (circles)  and $a_z^{\text{obs}}(R,z)$ (solid lines)  in function of $|z|$ for $R=10,  13, 16, 19$ kpc. 
Bottom panel: values of $\overline{h_z}$ for the DMD and NFW model that give the best agreement with the data together with the behaviors for the thin, thick and gas disks. 
 \label{DR3_Bosma_2}}
\end{figure}

The upper panel of Fig. \ref{DR3_Bosma_2} shows that the agreement between $a_z^{\text{obs}}(R,z)$ and $a_z^{\text{mod}}(R,z)$ is less good but still  reasonable. Given that the DM disk is the heavier one it is not surprising that  the vertical height giving the best accordance with the data is closer to the gaseous's one, i.e. $\overline{h_z(R)} \approx h_{z,g}$, than to the thick disk one: this is indeed the case (see the bottom panel of Fig. \ref{DR3_Bosma_2}).  In the discussion section below we report the $\chi^2$ values of the fits. 


\subsection{The NFW halo model}
\begin{figure}
    \centering
	\includegraphics[width=0.5\textwidth]{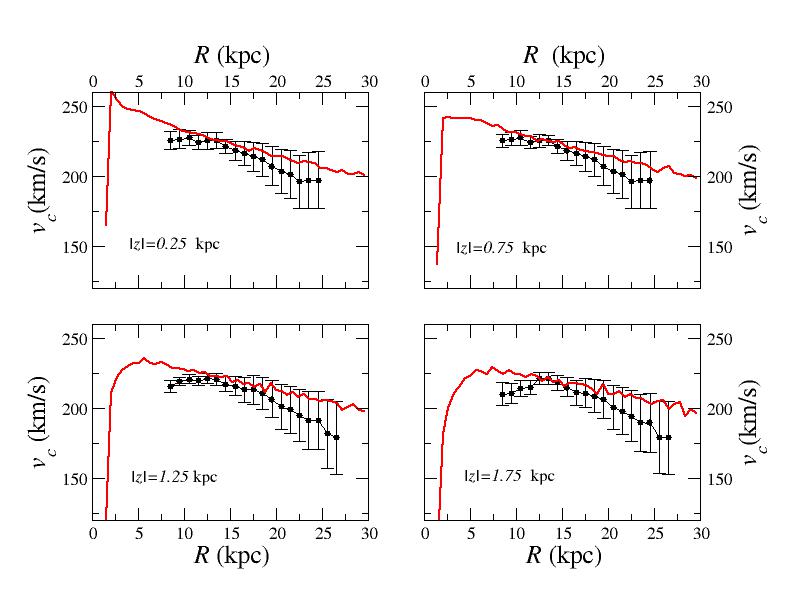}
   \caption{As Fig.\ref{DR3_Bosma} for the halo model. \label{DR3_halo_vc}}
\end{figure}

\begin{figure}
\includegraphics[width=0.5\textwidth]{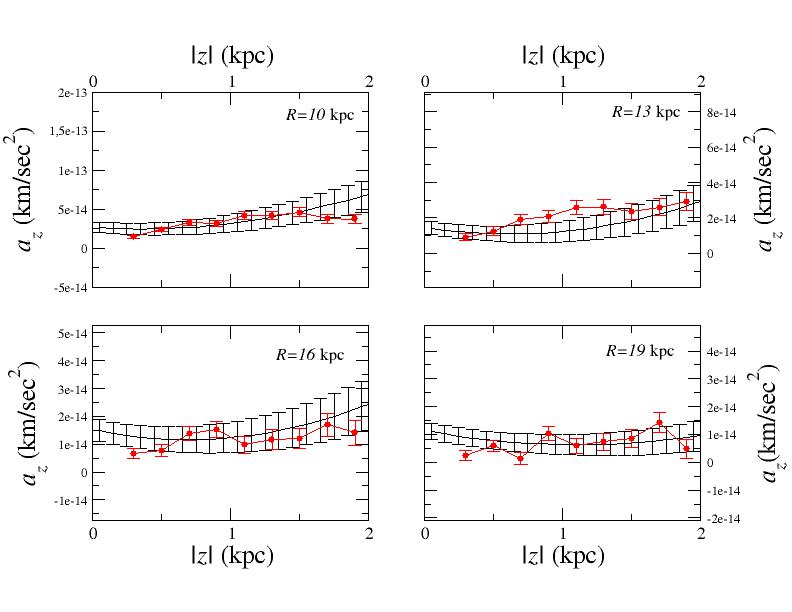}
   \caption{As the upper panel of Fig.\ref{DR3_Bosma_2} for the halo model. \label{DR3_halo_az}}
\end{figure}

For the case of the halo model the vertical density distribution in the disk depends solely on the baryonic components, as the free parameters in this case describe the NFW halo.  
{  Fig.\ref{DR3_halo_vc} }
shows the results for the case in which $\rho_0=9.4 \times 10^{-25}$ gr cm$^{-3}$ and $r_s=12.5$ kpc \citep{SylosLabini_etal_2023_MW}. In our study, we employ the same parameters as the fit without the flares, as the main mass component in this case arises from the spherical halo.  
The vertical accelerations $a_z^{\text{mod}}(R,z)$ have half the amplitude of those obtained in the DMD model. Thus, they are in agreement with $a_z^{\text{obs}}(R,z)$ only when assuming a higher value of $h_z(R)$ compared to the DMD model. This is consistent with the fact that, in this case, the gas disk has a negligible effect, while the actual $h_z(R)$ must be larger due to the contribution of the thick disk (see the bottom panel of 
{  Fig. \ref{DR3_halo_az})}.


\subsection{Discussion} 
In order to present a quantitive estimation of the relative performance of the DMD and NFW models, Tab.\ref{table2} provides the reduced $\chi^2$ values, with 2 degrees of freedom, for the fits of the generalized rotation curves in two models, considering four vertical slices. The fits were performed within the range of radial distances of 8.5-25 kpc (case [1]) and 13-25 kpc (case [2]). As mentioned above, the precise characterization of the different flares is needed to make an accurate fit at small radii. On the other hand, at large enough radial distances the generalized rotation curves are not affected by the specific features of the flares in the different mass components.  Comparing the generalized rotation curve $v_c(R,z)$ with the DMD model, we observe that it exhibits a weaker agreement with the model prediction for all four considered values of $z$ at large radial distance. Indeed,  the rotation curves decay slower than the Gaia DR3 data. As mentioned previously, to address this, one possible approach is to consider a generalized NFW profile or an Einasto profile, as demonstrated by \cite{Ou_etal_2024}.

\begin{table}
\begin{center}
\begin{tabular}{| c |c |c |c |c |   }
\hline 
 slice ($z$ kpc) 	&	$\chi^2_{\text{nfw}}$ [1] & $\chi^2_{\text{dmd}}$ [1] &	$\chi^2_{\text{nfw}}$ [2] & $\chi^2_{\text{dmd}}$ [2] \\ 
\hline
0.25 &	0.49&	1.0&	        0.40&	0.22 \\
0.75&	0.37&	0.37&	0.35&	0.10\\
1.25&	1.2&	      	0.34&	0.52&	0.11\\
1.75&	0.8&		0.6&	         0.48&	0.45\\
\hline
\end{tabular}
\end{center}
\caption{Reduced $\chi^2$ values for the fits of the generalized rotation curves in two models, considering four vertical slices. The fits were performed within the range of radial distances 8.5-25 kpc (case [1]) and 13-25 kpc (case [2]). }
\label{table2}
\end{table}

Similarly Tab.\ref{table3} provides the reduced $\chi^2$ values for the fits of the vertical acceleration in two models, considering four vertical slices. The fits were performed within the range of vertical heights $|z| \in (0,2)$ kpc. The
fits in this case are similar.
\begin{table}
\begin{center}
\begin{tabular}{| c | c | c |   }
\hline
 slice ($R$ kpc) 	&	$\chi^2_{\text{nfw}}$ & $\chi^2_{\text{dmd}}$ \\ 
\hline
10 &	0.41&	1.29 \\
13&	1.10&	0.58 \\
16&	0.52& 0.21\\
19&	1.06&	0.21\\
\hline
\end{tabular}
\end{center}
\caption{Reduced $\chi^2$ values for the fits of the vertical acceleration in two models, considering four vertical slices. The fits were performed within the range of vertical heights  0-2 kpc.}
\label{table3}
\end{table}

{  Finally { Tab.\ref{table4} }summarizes the values of the best fit parameters for the two mass models that we considered.} 
\begin{table}
\begin{center}
\begin{tabular}{ | c c | c c |}
\hline
$\rho_0$	                                        &	$r_s$                  &      $M_{\text{dm,disk}}$                          &  $h_{g,R}$ \\ 
\hline
$9.4 \times 10^{-25}$ gr cm$^{-3}$  &	$12.5$ kpc 	& 	$ 9.0 \times 10^{10} M_\odot$ &  5 kpc\\
\hline
\end{tabular}
\end{center}
\caption{ Best fit parameters for the two mass models considered.}
\label{table4}
\end{table}

\section{Conclusions}
\label{conclusion}

In this work, by using the data of the Gaia DR3 catalog analyzed by \cite{Wang_etal_2023}, who have measured the kinematic moments in the anti-center region, we have determined the generalized rotation curves $v_c(R,z)$ for radial distances in the range from 8.5 kpc to 25 kpc and vertical heights in the range from -2 kpc to 2 kpc. We have then used  $v_c(R,z)$ at different vertical heights to constrain the matter distribution in two distinct mass models: the first model adopts the canonical NFW halo model, while the second model, inspired by the Bosma effect, assumes that dark matter is confined to the Galactic plane and follows the distribution of neutral hydrogen. {  \cite{SylosLabini_etal_2023_MW} used both the observed profiles of \HI{} and \HI{}+H$_{2}$ for the fit, and we refer to that work for further details on this issue.}

Best-fitting the NFW model gives a virial mass $M_{\text{vir}} = (6.5 \pm 0.5) \times 10^{11} M_\odot$, whereas for the DMD model a total mass of $M_{\text{dmd}}= (1.71 \pm 0.2) \times 10^{11} M_\odot$. In our analysis, we have found that the DMD model generally provides a  better fit to the generalized rotation curves compared to the NFW model: this occurs especially at large radii, i.e. $R>13$ kpc where the rotation curves show a decay with radial distance. At small $R$, i.e. $R \le 12$ kpc where the rotation curve for vertical heights $|z| >1$ kpc show a rapid decline with $R$ the agreement is slightly better for the DMD than for the NFW: however to obtain more stringent constrains in this range of radial distances a precise characterization of the flares in the different mass components and a  full minimization procedure is necessary. This will be implemented in future works together with the study of modified gravity models. 

We conclude that examining the generalized rotation curves at different vertical heights allows us to gain insights into the vertical extent and shape of the DM component. If the DM distribution is primarily flattened within the galactic disk, we would expect to observe distinct variations at small radii (i.e., $R \approx 5-10$ kpc) in the rotation curves as the vertical distance changes. Conversely, if the DM distribution is more spherically symmetric, we would anticipate minimal changes in the rotation curves across different vertical heights. The analysis of the vertical acceleration does not give clear constraints on the different mass models, as an additional parameter, i.e., the vertical characteristic length, crucially determines its amplitude.

Overall, the forthcoming Gaia data release is indeed anticipated to provide a broader range of radial and vertical distance measurements with improved accuracy. This expanded dataset will be valuable for constraining the performance of models, particularly through the simultaneous fitting of generalized rotation curves within the 5-30 kpc range of radial distances and will probably allow to explore a larger range of vertical heights, i.e. $|z|<3$. These improved constraints have the potential to yield clearer conclusions regarding the geometric properties of the DM component.

\section*{Acknowledgments}
I warmly acknowledge the collaboration of Martin Lopez-Corredoira for his many discussions and constructive criticisms. Together, we developed the Lucy Inversion Method, and he also provided the Gaia DR3 sample used in this work.
I also thank Roberto Capuzzo Dolcetta, Michael Joyce and Antonio Tedesco for useful comments and discussions. 
I also acknowledge an anonymous referee for a detailed list of comments and suggestions. 
European Space Agency (ESA) mission Gaia 
(https://www.cosmos.esa.int/gaia), processed by the Gaia Data Processing and
Analysis Consortium (DPAC) 
Funding for the DPAC has been
provided by national institutions, in particular the institutions
participating in the Gaia Multilateral Agreement.

\end{document}